\documentclass[10pt]{iopart}
\pdfoutput=1
\usepackage{graphicx}
\usepackage{amssymb}
\usepackage{xcolor}
\usepackage{iopams} 

\eqnobysec

\newcommand{\beq}{\begin{equation}}
\newcommand{\beqa}{\begin{eqnarray}}
\newcommand{\eeq}{\end{equation}}
\newcommand{\eeqa}{\end{eqnarray}}

\newcommand{\zeq}[1]{\mathrel{\mathop{=}\limits_{#1}^{}}}
\def\stackunder#1#2{\mathrel{\mathop{#2}\limits_{#1}}}

\newcommand{\text}[1]{\hbox{#1}}

\newcommand{\prob}{\mathop{\rm Prob}\nolimits}

\def\sign{\mathop{\rm sign}}

\newcommand{\s}{{\sigma}}

\begin{document}
\date{\today}

\title[Freezing in stripe states for kinetic Ising models]{Freezing in stripe states for kinetic Ising models: a comparative study of three dynamics}

\date{\today}
\author{Claude Godr\`eche$^1$ and Michel Pleimling$^{2,3,4}$}
\address{$^1$Institut de Physique Th\'eorique, Universit\'e Paris-Saclay,
91191 Gif-sur-Yvette, France}\smallskip
\address{$^2$Department of Physics,
Virginia Polytechnic Institute and State University,
Blacksburg, Virginia 24061-0435, USA}\smallskip
\address{$^3$Center for Soft Matter and Biological Physics,
Virginia Polytechnic Institute and State University, 
Blacksburg, Virginia 24061-0435, USA}\smallskip
\address{$^4$Academy of Integrated Science,
Virginia Polytechnic Institute and State University, 
Blacksburg, Virginia 24061-0405, USA}\smallskip

\begin{abstract}
We present a comparative study of the fate of an Ising ferromagnet on the square lattice with periodic boundary conditions evolving under three different zero-temperature dynamics.
The first one is Glauber dynamics, the two other dynamics correspond to two limits of the directed Ising model,
defined by rules that break the full symmetry of the former, yet sharing the same Boltzmann-Gibbs distribution at stationarity.
In one of these limits the directed Ising model is reversible, in the other one it is irreversible.
For the kinetic Ising-Glauber model, several recent studies have demonstrated the role of critical percolation to predict the probabilities for the system to reach the ground state or to fall in a metastable state.
We investigate to what extent the predictions coming from critical percolation still apply to the two other dynamics.
\end{abstract}


%
\section{Introduction}

Determining how a system relaxes to equilibrium is a fundamental question of nonequilibrium statistical mechanics.
For instance, the question of how equilibrium is reached under reversible dynamics for a ferromagnetic spin system evolving from a random disordered initial condition has a long history, starting with the analytical treatment by Glauber of the one-dimensional kinetic Ising model \cite{glau}.
In two and three dimensions the dynamics of an Ising spin system quenched from a random disordered (high temperature) initial condition and relaxing to equilibrium under single-spin-flip Glauber dynamics at a finite temperature below criticality exhibits phase ordering \cite{bray,reviewgl,malte}.
The time taken by a finite system to reach this equilibrium state scales typically as the square of the system size.

The zero-temperature dynamics is special because, instead of reaching the ground state, the system can get trapped in infinitely long-lived metastable states.
Consider Ising spins $\s=\pm1$ with energy function (or Hamiltonian)
\beq\label{eq:Ising}
E=-J\sum_{(i,j)}\s_i\s_j,
\eeq
where $(i,j)$ are nearest-neighbours sites and $J>0$ is the coupling constant.
For zero-temperature Glauber dynamics,
moves such that $\Delta E>0$ have zero rates,
moves such that $\Delta E=0$ have rate $1/2$ and those leading to $\Delta E<0$ have rate $1$ (up to a scale of time).
Thus, for instance, on the square lattice, horizontal or vertical straight stripes of width larger than unity can not evolve under this zero-temperature dynamics.
For the square lattice with either periodic or open boundary conditions there appears to be a nonzero probability of reaching these stripe states \cite{spirin1,spirin2}, for which an exact value has been recently proposed \cite{barros,olejarz}.
The prediction relies on the unexpected relationship between the fate of the finite system at very long times, beyond the coarsening regime, and the analysis of the short-time domain structure of the coarsening system, using concepts of critical percolation \cite{barros,olejarz}.
According to \cite{olejarz} this result applies to any curvature-driven coarsening process with a non-conserved scalar order parameter.
Further studies in this direction are reviewed in \cite{picco}.

Here we make a comparative study of the fate of an Ising ferromagnet on the square lattice with periodic boundary conditions evolving from a random disordered initial condition under three different dynamics, in the limit of zero temperature.
The first one is Glauber dynamics.
The two other ones are defined by rules that break the full symmetry of the former.
They correspond to two limits of the two-dimensional directed (kinetic) Ising model \cite{gb2009,cg2011,cg2013,gp2014,gl2015,gp2015,gl2017}, whose definition, given below, involves an asymmetry parameter $V$.
The directed Ising model is itself an extension of the K\"unsch model \cite{kun} (see the appendix).
\begin{itemize}
\item For $V=0$ the model is anisotropic.
This is due to the fact that the influence on the central (flipping) spin comes from two distinct groups of neighbours separately, the North-East group on one hand and the South-West group on the other hand\footnote{There is nothing special about the North-East to South-West direction.
The model could alternatively be defined by privileging the North-West to South-East direction.}.
This is in contrast with what prevails for Glauber dynamics where all four neighbours of the flipping spin play the same role.
The dynamics is nevertheless reversible.

\item For $V=\pm1$ the dynamics is fully asymmetric: amongst the four neighbours of the flipping spin, only two of them have an influence on the central spin (North and East, if $V=1$, South and West if $V=-1$); this asymmetry implies the irreversibility of the dynamics.
\end{itemize}
The remarkable property of this dynamical model is that, for any value of the asymmetry parameter $V$, and at any finite temperature, its stationary state has the Gibbs-Boltzmann distribution for the Ising Hamiltonian (\ref{eq:Ising}), even though, as soon as $V\ne0$, the dynamics is irreversible \cite{gb2009,cg2013,gp2014}.
For $V=0$ the dynamics leads to equilibrium, for $V\ne0$ it leads to a nonequilibrium stationary state.

The questions raised in the present work are: what is the ultimate fate of these dynamics at zero temperature, in particular how rapidly and how often is the ground state attained, compared to the well-studied Glauber case?
Do the predictions of critical percolation still hold?
Are there differences in behaviour between the $V=0$ dynamics, which is reversible and obeys detailed balance, and the $V=1$ dynamics which is irreversible and violates detailed balance?

The present work is the natural sequel of \cite{gp2015}, which essentially focussed on the coarsening regime, i.e., long before the ultimate regime where the system reaches the ground state or falls into an infinitely long-lived metastable state.

\section{Rules of zero-temperature dynamics}
\label{sec:rules}
We consider a system of Ising spins on a square lattice of linear size $L$, with periodic boundary conditions,
evolving from a random disordered initial condition.
The energy of the system is given in (\ref{eq:Ising}).
At each instant of time, a spin, denoted by $\s$, is chosen at random and flipped with rate $w$.
The variation in energy due to a flip reads
\beq
\Delta E=2\sigma (\sigma_E+\sigma_N+\sigma_W+\sigma_S)=2\sigma h,
\eeq
with the notation $\{\s_E,\s_N,\s_W,\s_S\}$ for the East, North, etc., neighbours of the central spin $\s$, 
and where $h$ is the local field 
\beq
h=\s_E+\s_N+\s_W+\s_S.
\eeq

\subsection{Glauber rates}

For a given configuration of the spins, the zero-temperature Glauber rate reads
\beq\label{eq:glauberT0}
w=\frac{\alpha}{2}\left[1-\sigma\sign h\right],
\eeq
(with $\sign h=0$ if $h=0$) which implies that moves such that $\Delta E>0$ have zero rate,
moves such that $\Delta E=0$ (first line of figure \ref{fig:movesGlau}) have rate $1/2$ and those leading to $\Delta E<0$ (second and third lines of figure \ref{fig:movesGlau}) have rate $1$, in units of $\alpha$, as recalled in the introduction.
Thus, on the square lattice, horizontal or vertical straight stripes of width larger than unity can not evolve under this zero-temperature dynamics.

\subsection{Rates for the directed Ising model}
For the directed Ising model with general asymmetry parameter $V$, the zero-temperature rate is given by
\beq\label{eq:rateT0}
\fl w=\frac{\alpha}{2}
\left[\frac{1+V}{2}(1-\s\s_E)(1- \s\s_N)
+\frac{1-V}{2}(1- \s\s_W)(1- \s\s_S)\right].
\eeq
If $\s=+1$ (respectively $\s=-1$)
$w$ is non zero, either when the North and East spins are both negative (respectively positive) or when this holds for the South and West spins.
The corresponding configurations are depicted in figure \ref{fig:movesDIM}.
Hereafter we focus our attention on the two extreme cases $V=0$ and $V=1$ (the physical properties of the model with $V\ne0$ are qualitatively the same as when $V=1$).
\begin{itemize}
\item For the former case ($V=0$), the North-East and South-West groups of spins have equal influence on the flipping spin $\s$.
The corresponding rates are given in figure \ref{fig:movesDIM}: $w=1$ for the first and second lines, $w=2$ for the second line (in units of $\alpha$).
\item In the latter case ($V=1$), only the configurations where the North and East spins are both negative (respectively positive) allow the central positive (respectively negative) spin $\s$ to flip.
The corresponding rate is equal to 2, in units of $\alpha$.

\end{itemize}

\begin{figure}[ht]
\begin{center}
\includegraphics[angle=0,width=0.9\linewidth]{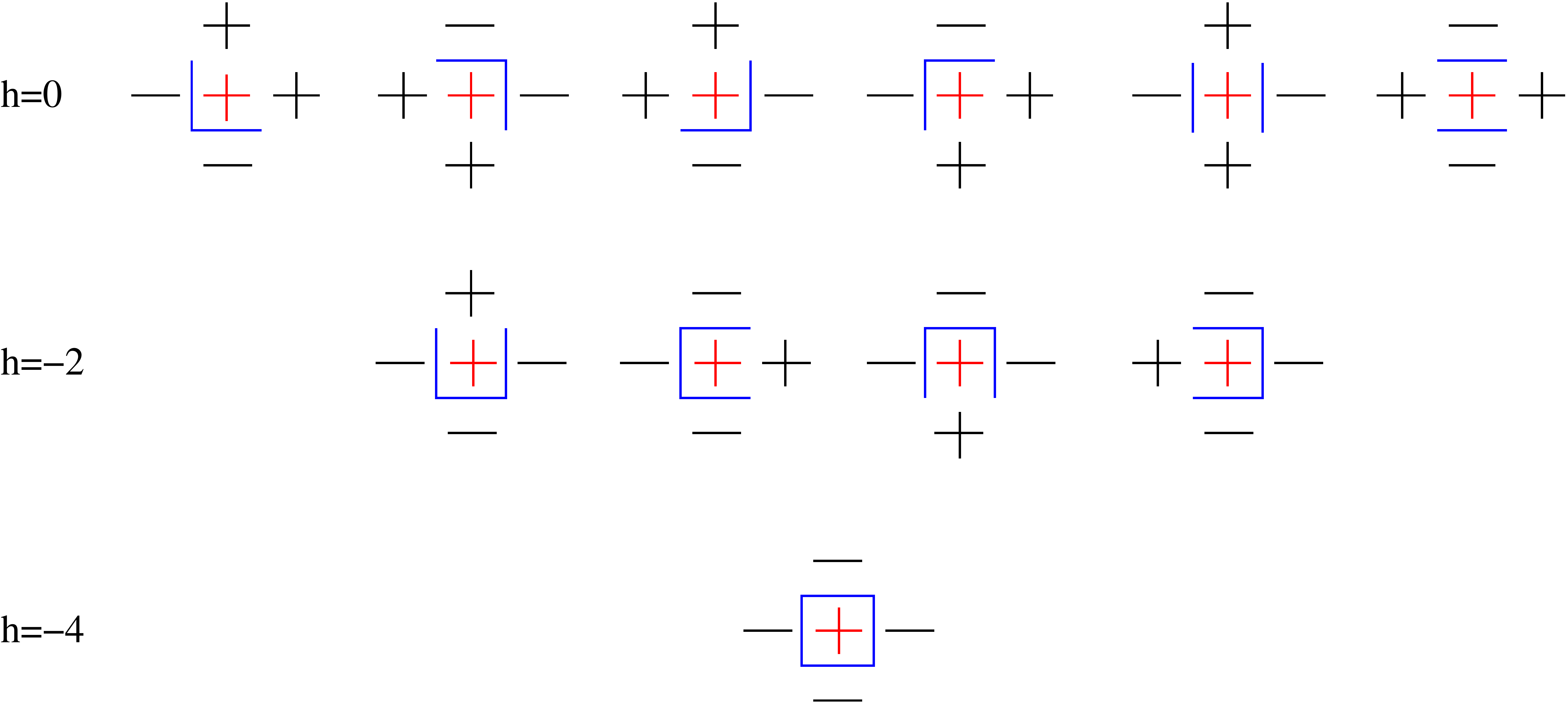}
\end{center}
\caption{Local configurations with non-vanishing values of the flipping rate for the central spin (in red)
for zero-temperature Glauber dynamics (\ref{eq:glauberT0}).
The blue segments represent the domain walls.
The first line corresponds to a value of the local field $h=0$, the second one to $h=-2$, the third one to $h=-4$.
The associated values of the rate are, in units of $\alpha$,
first line: $1/2$;
second line: $1$;
third line: $1$.
}
\label{fig:movesGlau}
\end{figure}

\begin{figure}[ht]
\begin{center}
\includegraphics[angle=0,width=0.6\linewidth]{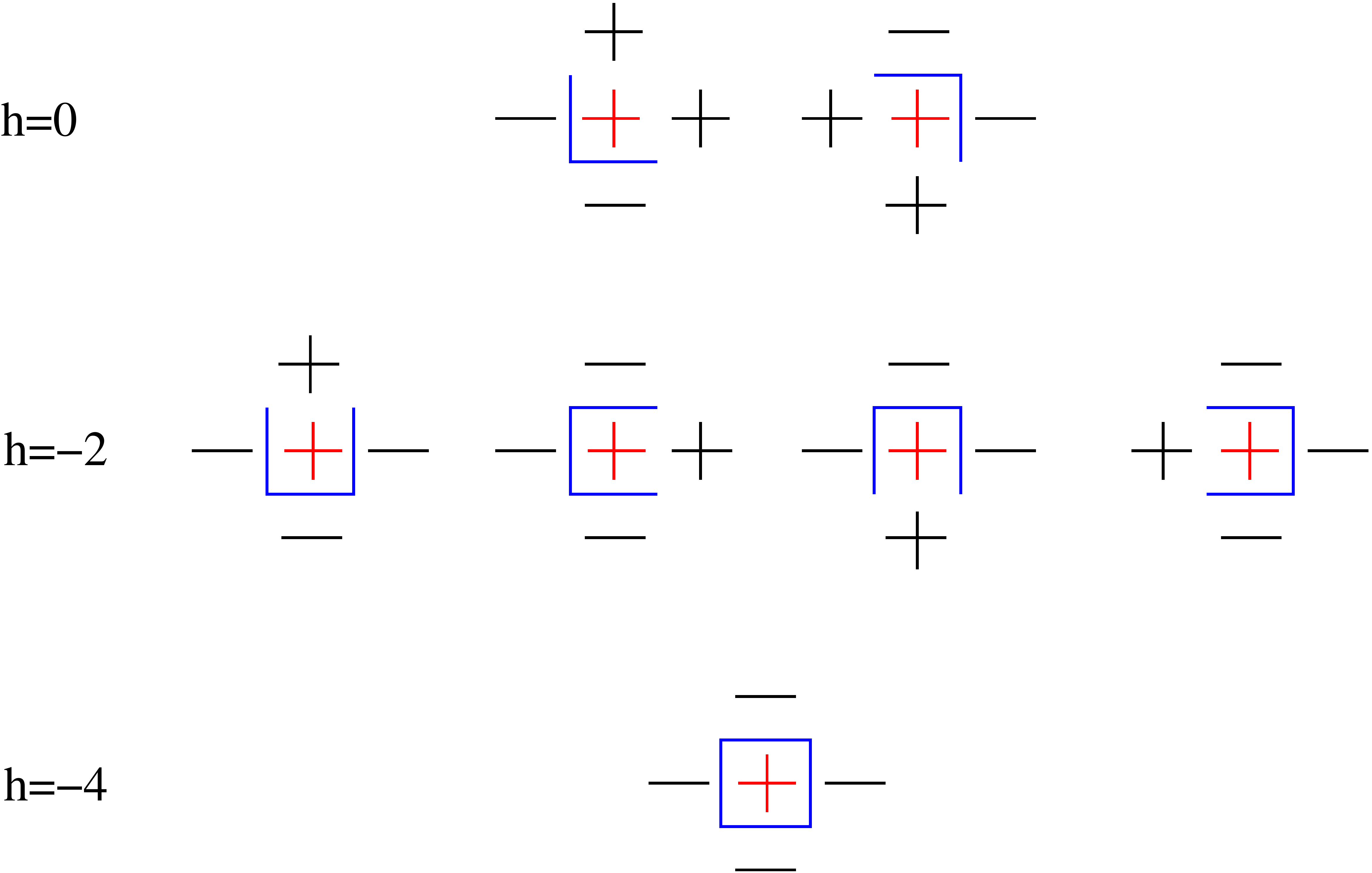}
\end{center}
\caption{Local configurations with non-vanishing flipping rates for the central spin (in red)
in the zero-temperature directed Ising model (\ref{eq:rateT0}).
The associated values of the rate are, from left to right, in units of $\alpha$,
first line: $1-V$, $1+V$;
second line: $1-V$, $1-V$, $1+V$, $1+V$;
third line: $2$.
The 4 configurations in the first line ($h=0$) of figure \ref{fig:movesGlau} which are not represented here have zero rate.
}
\label{fig:movesDIM}
\end{figure}

Due to the spin-reversal symmetry of the expressions (\ref{eq:glauberT0}) and (\ref{eq:rateT0}), the rates in figures \ref{fig:movesGlau} and \ref{fig:movesDIM} are unchanged if one reverts the signs of all spins.

As shown on figures \ref{fig:movesGlau} and \ref{fig:movesDIM}, the freedom for a spin to flip is restrained when passing successively from Glauber dynamics to $V=0$ dynamics, then to $V=1$ dynamics, since the number of `flat' directions in the energy landscape ($\Delta E=0$) decreases when passing from one model to the next one.
Moreover, for $V=1$ the number of allowed directions of descent with $\Delta E<0$ is also reduced compared to the two other dynamics.

Note the existence of horizontal (vertical) stripe states made of one line (column) of minority spins for the directed Ising model, while for Glauber dynamics these stripes have at least a width equal to 2.

\begin{figure}[ht]
\begin{center}
\includegraphics[angle=0,width=0.4\linewidth]{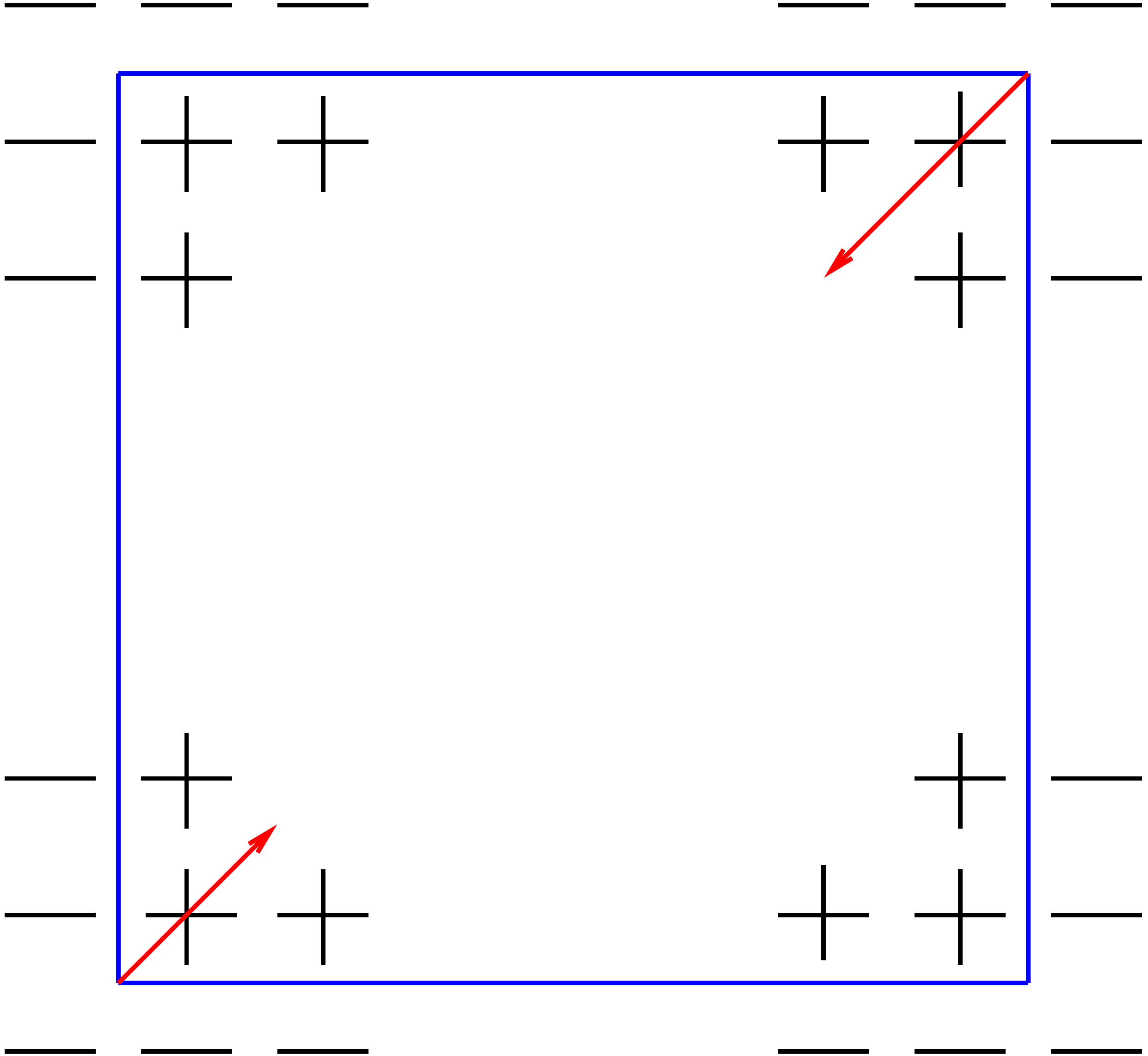}
\end{center}
\caption{Possible moves for a minority ($+$) square for the zero-temperature directed Ising model (\ref{eq:rateT0}).
Only the North-East and South-West corners move.
The big arrow corresponds to a rate equal to $1+V$ (in units of $\alpha$), the smaller one to $1-V$ (see the first line of figure \ref{fig:movesDIM}).
The same holds for a minority ($-$) square.
For Glauber dynamics the flipping rate is the same for the four corners.
}
\label{fig:moves2}
\end{figure}

\begin{figure}[ht]
\begin{center}
\includegraphics[angle=0,width=0.7\linewidth]{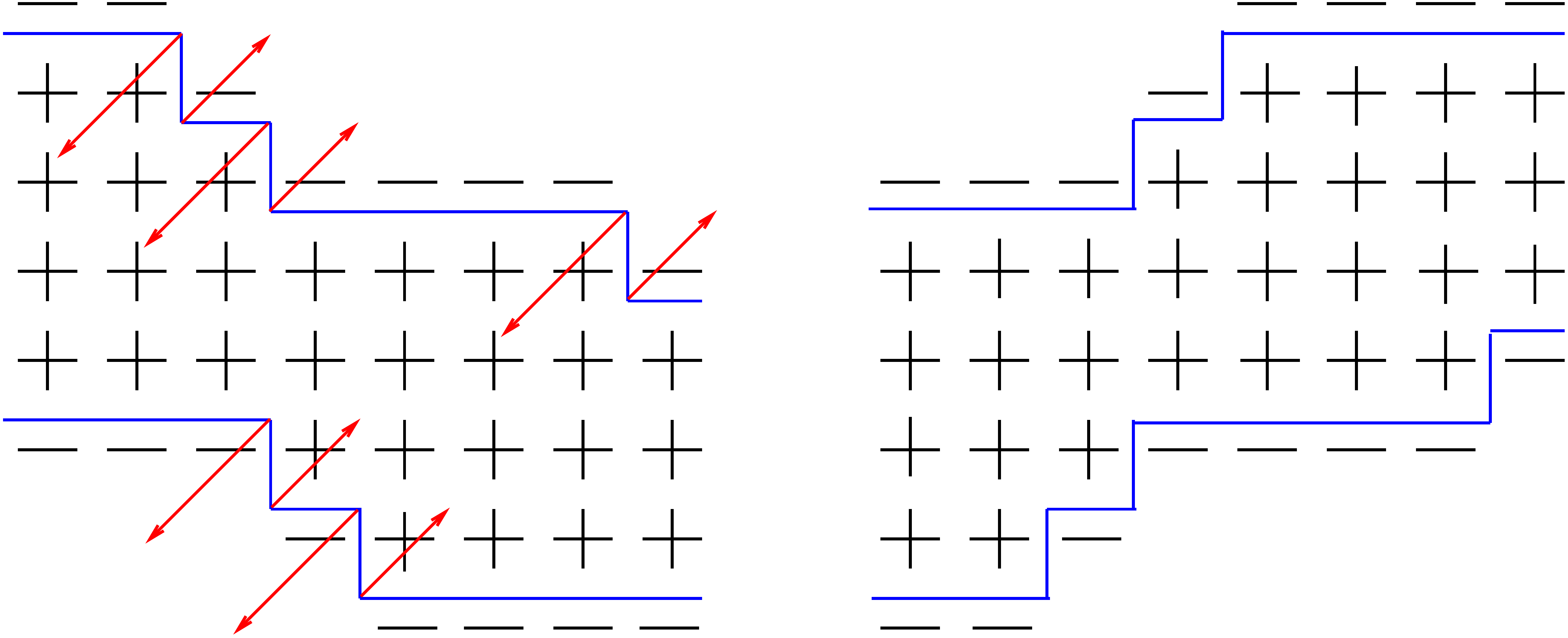}
\end{center}
\caption{Fate of a descending or an ascending stripe under the zero-temperature dynamics of the directed Ising model (\ref{eq:rateT0}).
Left: a descending stripe is unstable.
Right: an ascending stripe is blocked.
For Glauber dynamics both stripes are unstable. 
}
\label{fig:moves3}
\end{figure}

Figure \ref{fig:moves2} depicts the possible moves of a minority $(+)$ square under the dynamics (\ref{eq:rateT0}) with a generic value of the asymmetry parameter $V$.
Only the North-East and South-West corners can move, as can be seen from the first line of figure \ref{fig:movesDIM}.
We thus infer that an ascending stripe (oriented in the North-East direction) will not move, even when $V=0$, whereas a descending stripe will, as depicted in figure \ref{fig:moves3}.
This figure illustrates the anisotropy of the dynamics of the directed Ising model, even if $V=0$.
(Examples of such configurations are given in \cite{gp2015}, see figures 16 and 17 therein.)

We refer the reader to the appendix for a more complete account of the dynamical rules defining the directed Ising model, valid at any finite temperature.
Equations (\ref{eq:glauberT0}) and (\ref{eq:rateT0}) can be obtained from (\ref{glau2D}) and (\ref{eq:rate}) by setting 
$\gamma=1$.


%
%
\section{Theoretical predictions for critical percolation on the torus}
\label{sec:theory}

\subsection{Percolation clusters}

A summary of formulas useful for the sequel, taken from \cite{moloney} (see also \cite{pinson,barros,olejarz}), is as follows.
Consider a rectangular lattice with periodic boundary conditions and with aspect ratio 
\beq
r=\frac{L_y}{L_x}.
\eeq
We are interested in paths winding around the lattice:
$(a,b)$ count the number of windings in the horizontal and vertical directions.
The conventions of \cite{moloney} are as follows:
\begin{itemize}
\item If $a=0$ then the only path which is non homotopic to a point and which does not intersect itself is such that $b=1$.
\item If $a\ne0$ and $b\ne0$ then $a$ and $b$ must be relatively prime if the path does not intersect itself.
\item One takes $a\ge0$.
\item The path with cross topology is produced by intersecting  $(1,0)$ and $(0,1)$.
\end{itemize}

The following paragraph is taken verbatim from \cite{moloney}.
`To transfer the notion of winding from paths to clusters, one considers all (topologically different) paths within a cluster.
If there is a path with non-zero winding numbers that crosses only paths which are homotopic to a point or have the same winding numbers, the cluster is assigned the winding numbers of the path. 
If there is no winding path at all, then this cluster itself is said to be homotopic to a point.
It transpires that in any other case the cluster contains a cross-topological path and the cluster is then said to have a cross topology itself.'

In the limit $L_x$ and $L_y$ $\to\infty$ the probability for a spanning cluster, in continuum percolation, with winding numbers $(a,b)$ on this rectangle is given by
\beq\label{eq:pab}
 \pi_{a,b}(r)=\frac{Z_{a,b}(6,r)-2Z_{a,b}(8/3,r)+Z_{a,b}(2/3,r)}{2[\eta(\e^{-2\pi r})]^2},
\eeq
where
\beq
Z_{a,b}(g,r)=\sqrt{\frac{g}{r}}
\sum_{j=-\infty}^{\infty}\e^{-\pi g(a^2/r+b^2 r)j^2},
\eeq
and $\eta$ is the Dedekind function
\beq
\eta(q)=q^{1/24}\prod_{k\ge1}(1-q^k).
\eeq
Note that $ \pi_{a,b}(r)= \pi_{b,a}(1/r)$.
For instance, $ \pi_{1,0}$ is the probability for (one or several) horizontal stripes (also denoted by $\pi_h$), $ \pi_{0,1}$ the probability for (one or several) vertical stripes (also denoted by $\pi_v$), $ \pi_{1,1}$ the probability of a stripe in the $(1,1)$ direction, and so on.

For the cross topology the probability is
\beqa\label{eq:pX}
\pi_{X}(r)=\frac{1}{[\eta(\e^{-2\pi r})]^2}\sqrt{\frac{3r}{8}}
\Big[&&\frac{1}{2}Z_{1,0}(8/3,r)Z_{0,1}(8/3,r)
\nonumber\\
&&-Z_{1,0}(2/3,r)Z_{0,1}(2/3,r)\Big].
\eeqa
The probability that all clusters in a given configuration are homotopic to a point is denoted by $\pi_{0}(r)$ and equal to  $\pi_{X}(r)$.

\subsection{Ising clusters}
We now consider percolation on Ising clusters.
We denote by $ \pi_{h,v}= \pi_{X}+\pi_{0}$ the probability for a cluster percolating in the two directions and by $\pi_d$ the probability for diagonal stripes,
\beq
 \pi_d= \pi_{1,1}+\pi_{1,-1} +\pi_{1,2}+\pi_{1,-2}+\pi_{2,1}+\pi_{2,-1}+\pi_{1,3}+\cdots.
 \eeq
The latter can be decomposed into two components, corresponding respectively to ascending or descending stripes,
 \beq
 \pi_{d}=\stackunder{\pi_{asc}}{\underbrace{\pi_{1,1}+\pi_{1,2}+\pi_{2,1}+\cdots}}
 +\stackunder{\pi_{desc}}{\underbrace{\pi_{1,-1}+\pi_{1,-2}+\pi_{2,-1}+\cdots}},
 \eeq
The probabilities defined above satisfy the sum rule
 \beq
\pi_{h,v}+ \pi_{0,1}+\pi_{1,0}+\pi_{d}=1.
 \eeq

These definitions can be illustrated on the case of a square $L\times L$ system ($r=1$).
We have (using Mathematica and keeping the first digits only)
\beq\label{eq:p01}
 \pi_{0,1}= \pi_{1,0}=0.169415,
\eeq
\beq\label{eq:p11}
 \pi_{1,1}= \pi_{1,-1}=0.020979,
\eeq
\beq
\pi_{1,\pm2}=\pi_{2,\pm1}=0.000039,
\eeq
\beq
 \pi_{h,v}=0.619053,
\eeq
\beq\label{eq:pasc}
 \pi_{asc}=\pi_{desc}\approx \pi_{1,1}+\pi_{1,2}+\pi_{2,1}= 0.021058.
 \eeq

\begin{table}[ht]
\caption{Probabilities for blocked configurations with $r=1$ obtained by simulations
compared to the predictions of percolation (see (\ref{eq:p01}), (\ref{eq:pasc}) and (\ref{eq:pgG})).
While the agreement is very good for Glauber dynamics, discrepancies exist for the directed Ising model.}
\label{tab:resultat}
\begin{center}
\begin{tabular}{|c|c|c|c||c|c|}
\hline
dynamics&Glauber&$V=0$&$V=1$&percolation&\\
\hline
$p_{0,1}=p_{1,0}$&0.1695&0.1661&0.1660&$\pi_{0,1}=\pi_{1,0}$&0.16941\\
$p_{asc}$&0&0.0378&0.0378&$\pi_{asc}$&0.02106\\
$p_{desc}$&0&0&0&$\pi_{desc}$&0.02106\\
$p_{g}$&0.6609&&&$1-2\pi_{0,1}$&0.66117\\
$p_{g}$&&0.6303&0.6305&$1-2\pi_{0,1}-\pi_{asc}$&0.64011\\
\hline
\end{tabular}
\end{center}
\end{table}

\section{Probabilities to end in a given final state}
\label{sec:blocking}
 
For the three dynamics under study we denote the probabilities for the system evolving from a random disordered initial condition to end in a final state with the geometries
defined above by $p_{0,1}$, $p_{1,0}$, $p_{1,1}$, \dots,
and the probability of reaching the ground state by $p_g$.

For Glauber dynamics, as shown in \cite{barros,olejarz,picco} and confirmed by the present study, the predictions of critical percolation hold, with
\beqa
p_{0,1}=\pi_{0,1},\quad p_{1,0}=\pi_{1,0}.
\eeqa
Since the diagonal stripes are unstable by the dynamics we have $p_d=0$, and therefore $\pi_d$ contributes to the ground state, so
\beq\label{eq:sumrule}
p_g=\pi_{h,v} +\pi_d=1-(\pi_{0,1}+ \pi_{1,0}).
\eeq
For instance for $r=1$, this prediction yields
\beq\label{eq:pgG}
p_g = 0.66117,
\eeq
which is satisfied by numerical simulations, see table \ref{tab:resultat}.

For the directed Ising model, the question is whether the predictions of critical percolation hold.
Since the descending stripes are unstable by the dynamics ($p_{desc}=0$) the following sum rule holds,
\beq\label{eq:pgDIM}
p_g=
1-(p_{0,1}+ p_{1,0}+p_{asc}).
\eeq

\begin{figure}[h]
\begin{center}
\includegraphics[angle=0,width=0.7\linewidth]{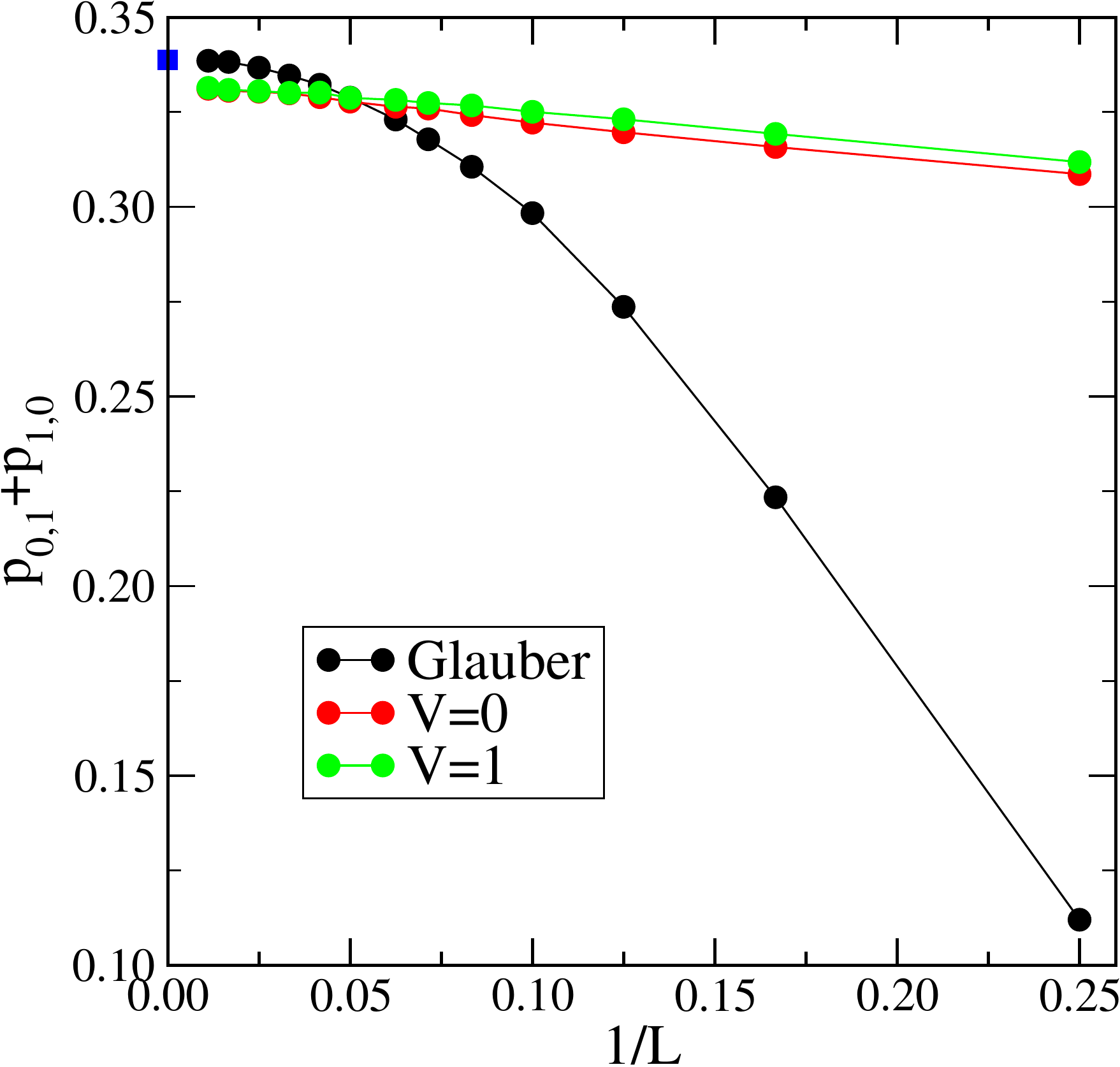}
\end{center}
\caption{
Probability $ p_{0,1}+ p_{1,0}$ that a square system with $L\times L$ spins ends in a vertical or horizontal stripe state. 
The blue square indicates the theoretical predication from percolation ($\approx0.339$), see table \ref{tab:resultat}.
In black, Glauber dynamics; in red, $V=0$; in green, $V=1$.
The data result from averaging over 6.4 million histories.
}
\label{fig:P10P01}
\end{figure}

The aim of this section is to give results of simulations on the three dynamics under study and compare them to the predictions of critical percolation.
In practice, in order to determine whether, for a given history, the system has reached its final state, we check after every Monte Carlo step whether there are still spins in the system that can flip.
When all activity has stopped, i.e., when the final state has been reached, we determine the nature of this final state
by identifying straight interfaces separating domains of positive and negative magnetization for vertical and horizontal stripe states,
or boundaries that are not straight for ascending stripe states.

\subsection{Results for aspect ratio $r=1$ and their size dependence}
\label{sec:extrapol}

We start by discussing the results of simulations on square systems ($r=1$) for the three dynamics.
Figure \ref{fig:P10P01} depicts the probability 
($ p_{0,1}+ p_{1,0}$) for the system to end in a (vertical or horizontal) stripe state
for sizes ranging from $L=4$ to $L=90$.

The blue square in this figure indicates the prediction of critical percolation ($\approx0.339$), see table \ref{tab:resultat}. 
The convergence to this predicted value is very good for Glauber dynamics.
For the directed Ising model with $V=0$ or $V=1$ the data seem to converge to a common value a few percent smaller ($\approx0.332$), see table \ref{tab:resultat}.
Note the difference in the size dependences of these data: while for the directed Ising model the behaviour seems linear, at least for larger systems, this is not the case for Glauber dynamics.

\begin{figure}[h]
\begin{center}
\includegraphics[angle=0,width=0.7\linewidth]{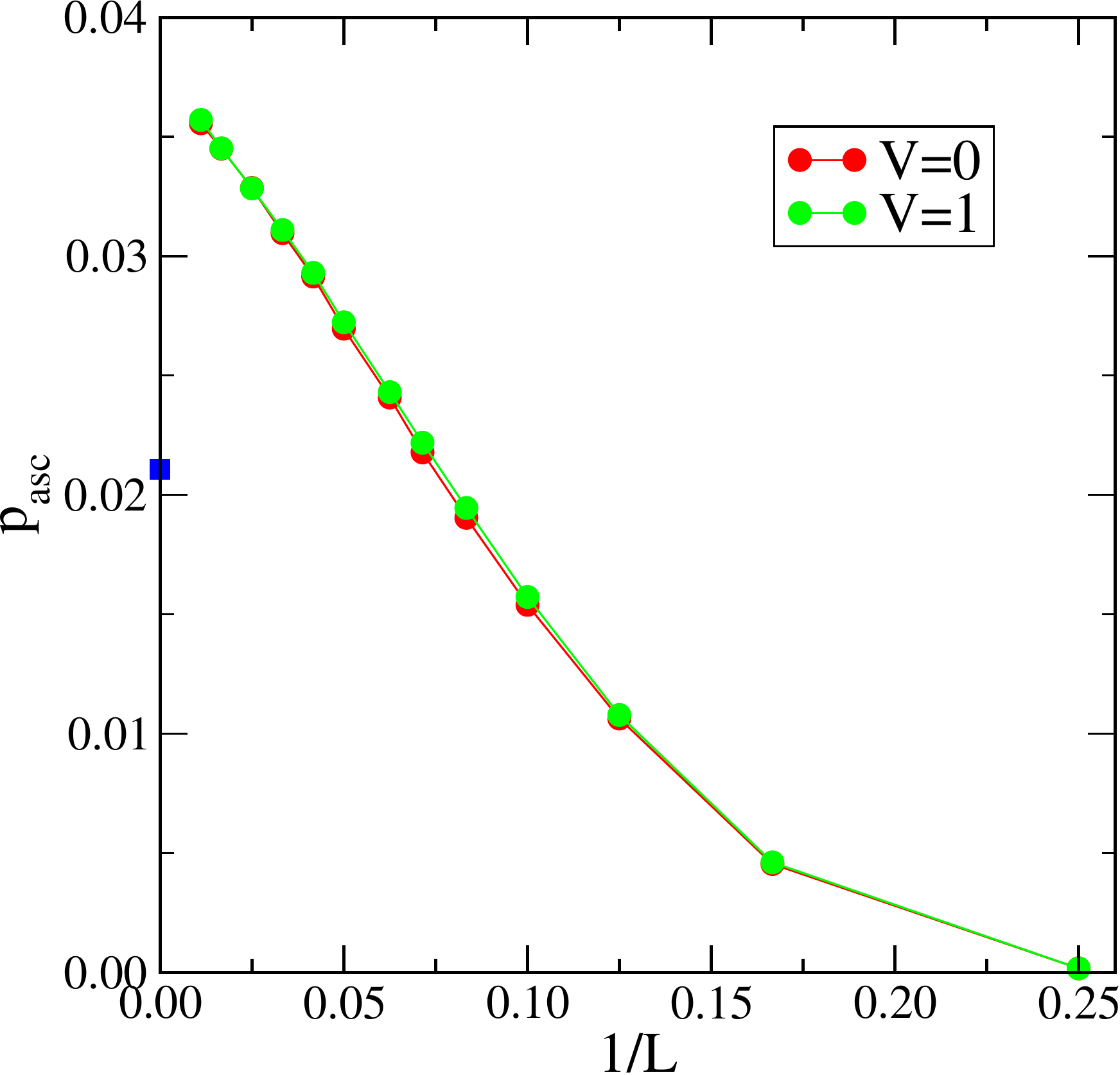}
\end{center}
\caption{
Probability $p_{asc}$ that a square system with $L\times L$ spins ends in an ascending stripe state. 
In red, $V=0$; in green, $V=1$.
The extrapolated value is far away from the prediction of critical percolation ($\approx 0.02106$, blue square), see table \ref{tab:resultat}.
}
\label{fig:Pd}
\end{figure}

Figure \ref{fig:Pd} depicts the probability $p_{asc}$ that a square system ends in an ascending stripe state for the directed Ising model with $V=0$ or $V=1$. 
The extrapolated value of this probability ($\approx 0.0378$) is far away from the prediction of critical percolation ($\approx 0.02106$), see table \ref{tab:resultat}.

\begin{figure}[h]
\begin{center}
\includegraphics[angle=0,width=0.7\linewidth]{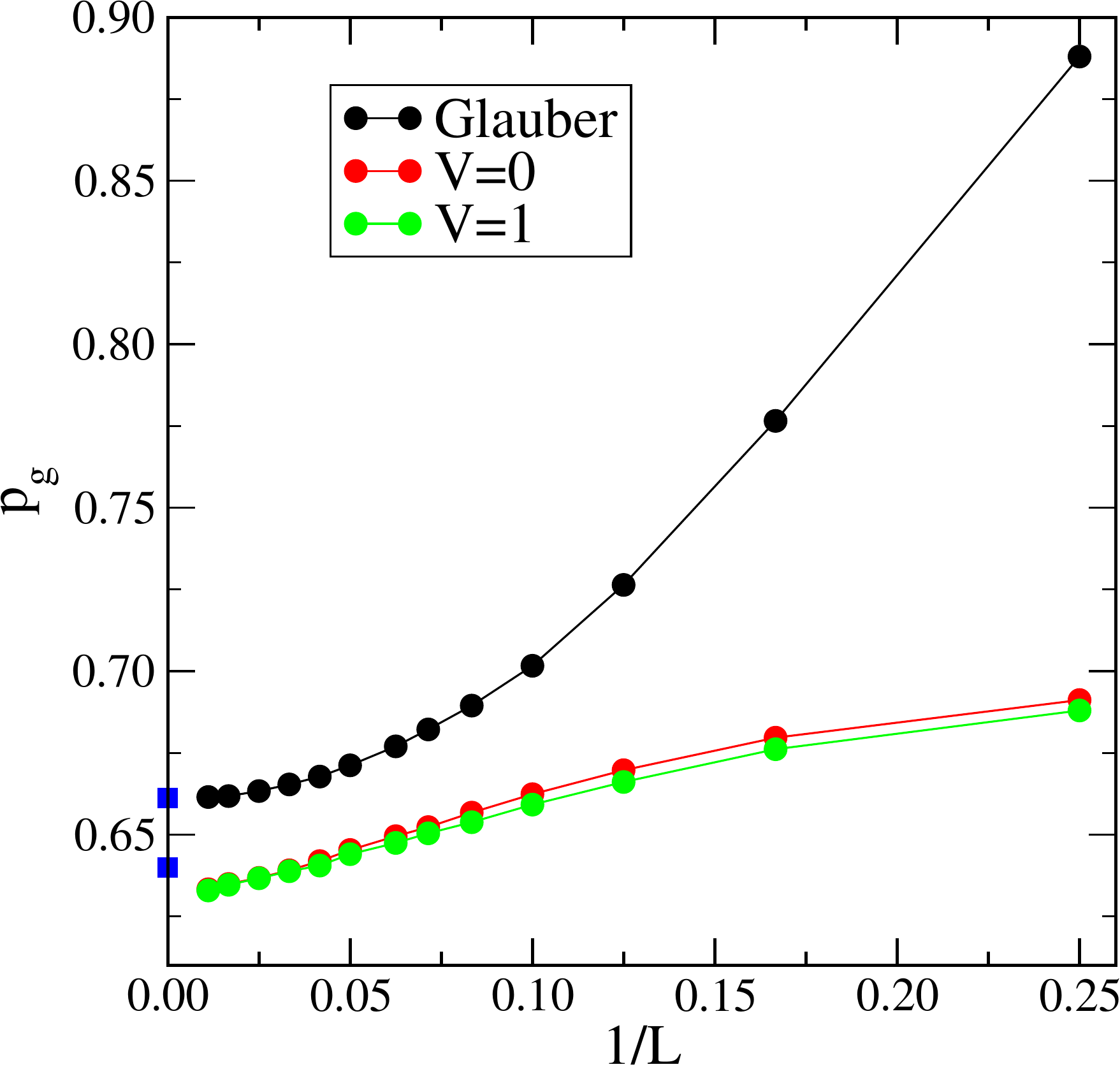}
\end{center}
\caption{
Probability $ p_{g}$ that a square system with $L\times L$ spins ends in the ground state. 
In black, Glauber dynamics; in red, $V=0$; in green, $V=1$.
The blue squares indicate the theoretical predications from percolation for Glauber dynamics ($\approx0.661$), and for the directed Ising model ($\approx 0.640$), see table \ref{tab:resultat}.
}
\label{fig:Pg}
\end{figure}

Finally figure \ref{fig:Pg} gives the probability for the system to reach the ground state, as inferred from the two previous figures.
This figure confirms the existence of discrepancies between the results of simulations for the directed Ising model and the predictions of critical percolation.


%
\subsection{Dependence in the aspect ratio}

\begin{figure}[h]
\begin{center}
\includegraphics[angle=0,width=0.7\linewidth]{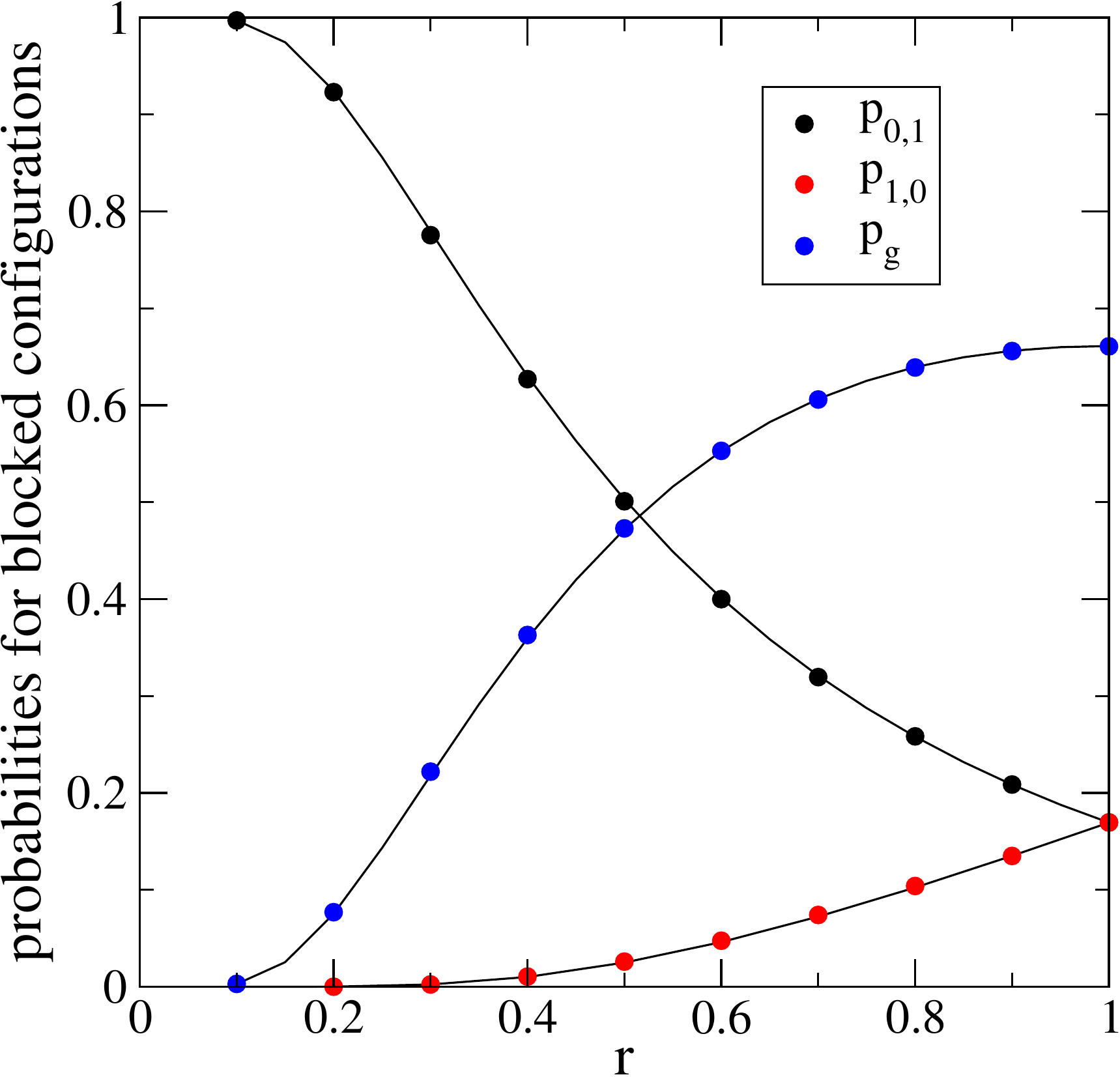} 
\end{center}
\caption{Probabilities for the system with Glauber dynamics 
to end in a vertical stripe state ($ p_{0,1}$),
an horizontal stripe state ($ p_{1,0}$), or in the ground state ($ p_{g}$).
The continuous lines are the theoretical predictions of percolation. 
The circles correspond to the numerically determined values using the same extrapolation scheme as in subsection 
\ref{sec:extrapol}.
}
\label{fig:probas}
\end{figure}

\begin{figure}[h]
\begin{center}
\includegraphics[angle=0,width=0.7\linewidth]{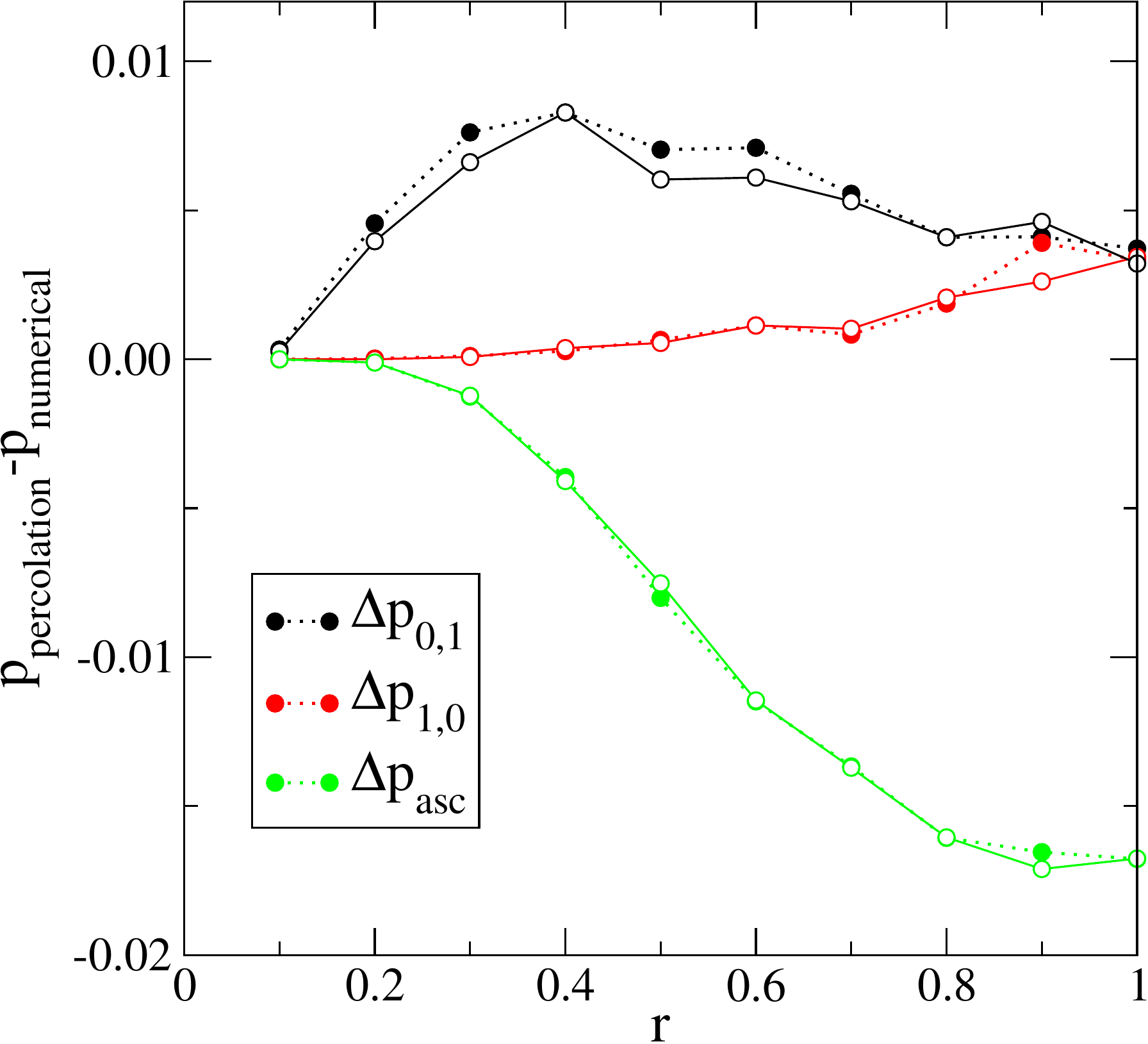}
\end{center}
\caption{
Differences between the theoretical predictions of percolation and the
numerically determined probabilities to end in a vertical stripe state ($ p_{0,1}$),
an horizontal stripe state ($ p_{1,0}$) or 
an ascending stripe state ($ p_{asc}$),  for the directed Ising model with $V=0$ (full circles) and $V=1$ (empty circles). 
The lines are guides to the eye.
The differences for $p_g$ are inferred from the previous ones (see (\ref{eq:pgDIM})).
The numerical values were obtained using the same extrapolation scheme as in subsection 
\ref{sec:extrapol}.
}
\label{fig:probas2}
\end{figure}

The same extrapolation scheme was applied to other values of $r$.
Figure \ref{fig:probas} depicts the probabilities for the system to end in a vertical stripe state ($ p_{0,1}$),
in an horizontal stripe state ($ p_{1,0}$),
or in the ground state ($ p_{g}$) for Glauber dynamics.
Figure \ref{fig:probas2} depicts, for the directed Ising model with $V=0$ and $V=1$, 
the difference between theoretical predictions coming from critical percolation and the
numerically determined probabilities to end in one of these three states as well as
in an ascending stripe state ($p_{asc}$)%
\footnote{For the values of $L$ and $r$ considered, only $(1,1)$, $(1,2)$, $(2,1)$ ascending stripes were observed.
Together the $(1,2)$ and $(2,1)$ stripes count for less than 0.6\% of all the ascending stripes.}.

The considered systems are of rectangular shape, with $L \times r \, L$ sites (and $0<r<1$). 
The number of sites in the horizontal direction, $L$, ranges from 20 to 160,
with the exception of $r=0.1$ for which $L$ ranges from 40 to 320.
For low values of $r$ one observes a proliferation of vertical stripes (up to nine stripes for the sizes considered). 
On the other hand, no states with more than two horizontal stripes were observed. 
The data for $V=0$ and $V=1$ are very close in most cases, as can be seen by the fact that the symbols for $V=0$ (full circles) often completely cover those for $V=1$ (empty circles). 
Simulations are run until cessation of any activity. 
The data result from averaging over 3.2 million independent histories (6.4 million histories for $r=1$).

For Glauber dynamics, the agreement of the simulation results with the theoretical predictions of percolation observed for $r=1$ (see table \ref{tab:resultat}) extends to smaller values of $r$.
This agreement confirms the observations made in \cite{barros,olejarz}.
On the other hand the discrepancies of the simulation results with the theoretical predictions of percolation observed for $r=1$, for the directed Ising model with $V=0$ or $V=1$, also extend to smaller values of $r$.
In relative value the discrepancies for $p_{0,1}$ and $p_{1,0}$ are small, while they are large for $p_{asc}$, as already observed for $r=1$.

\section{Hitting times}

The simulations of the previous section gave the probabilities, for the three dynamics, of reaching a given final state:
for Glauber dynamics the final state can be either the ground state or a (horizontal or vertical)
stripe state; for the directed Ising model the final state can alternatively be an ascending stripe state.
During the same simulations one can also monitor the time needed to reach this final state, or {\it hitting time} of this state.

\begin{table}[!h]
\caption{Median $\tau_g$ of the distribution of hitting times $t_g$ of the ground state.}
\label{tab:median_g}
\begin{center}
\begin{tabular}{|c|c|c|c|c|}
\hline 
& & Glauber & $V=0$ & $V=1$ \\ \hline
$r=1$ & $L=20$ & 150 & 149 & 58 \\ \hline
& $L=40$ & 630 & 621 & 133 \\ \hline
& $L=80$ & 2586 & 2537 & 291 \\ \hline
& $L=160$ & 10482 & 10269 & 613 \\ \hline
$r=0.5$ & $L=20$ & 60 & 64 & 34 \\ \hline 
& $L=40$ & 270 & 274 & 85 \\ \hline 
& $L=80$ & 1144 & 1141 & 193 \\ \hline 
& $L=160$ & 4702 & 4653 & 414 \\ \hline 
$r=0.1$ & $L=40$ & 16 & 17 & 14 \\ \hline
& $L=80$ & 66 & 80 & 42 \\ \hline
& $L=160$ & 316 & 350 & 107 \\ \hline
& $L=320$ & 1404 & 1502 & 246 \\ 
\hline
\end{tabular}
\end{center}
\end{table}

\begin{table}[!h]
\caption{Median $\tau_{v,1}$ of the distribution of hitting times $t_{v,1}$ of a single vertical stripe state.}
\label{tab:median_10} 
\begin{center}
\begin{tabular}{|c|c|c|c|c|}
\hline 
& & Glauber & $V=0$ & $V=1$ \\ \hline
$r=1$ & $L=20$ & 246 & 280 & 47 \\ \hline
& $L=40$ & 1132 & 1361 & 109 \\ \hline
& $L=80$ & 5072 & 6401 & 241 \\ \hline
& $L=160$ & 22204 & 29380 & 512 \\ \hline
$r=0.5$ & $L=20$ & 72 & 77 & 27 \\ \hline 
& $L=40$ & 330 & 376 & 67 \\ \hline 
& $L=80$ & 1480 & 1762 & 153 \\ \hline 
& $L=160$ & 6470 & 8023 & 332 \\ \hline 
$r=0.1$ & $L=40$ & 16 & 17 & 13 \\ \hline
& $L=80$ & 68 & 80 & 39 \\ \hline
& $L=160$ & 324 & 362 & 97 \\ \hline
& $L=320$ & 1436 & 1588 & 223 \\ 
\hline
\end{tabular}
\end{center}
\end{table}
%

\begin{table}[!h]
\caption{Median $\tau_{asc}$ of the distribution of hitting times $t_{asc}$ of an ascending stripe state.
No data are given for $r=0.1$ as no such state was encountered for these very asymmetric samples.}
\label{tab:median_asc}
\begin{center}
\begin{tabular}{|c|c|c|c|}
\hline
& & $V=0$ & $V=1$ \\ \hline
$r=1$ & $L=20$ & 62 & 24 \\ \hline
& $L=40$ & 316 & 61 \\ \hline
& $L=80$ & 1531 & 143 \\ \hline
& $L=160$ & 7073 & 313 \\ \hline
$r=0.5$ & $L=20$ & 191 & 41 \\ \hline
& $L=40$ & 988 & 100 \\ \hline
& $L=80$ & 4665 & 225 \\ \hline
& $L=160$ & 22655 & 484 \\ \hline
\end{tabular}
\end{center}
\end{table}

\subsection{Typical hitting times}

Let us first consider the hitting time $t_g$ of the ground state.
As will be seen in the next subsection the distribution of $t_g$ is bimodal.
However the weight of the contribution associated to the second mode is negligible compared to that of the first one.
We therefore take as a first measure of the typical hitting time the median $\tau_g$ of the whole distribution\footnote{See \cite{gp2015} for a similar analysis.}.
A more refined analysis is given in the next subsection.

The size and dynamics dependence of $\tau_g$ is illustrated in table \ref{tab:median_g}.
Table \ref{tab:median_10} provides similar results for the median $\tau_{v,1}$ of the distribution of the hitting time $t_{v,1}$ of a single vertical stripe state. 
Finally, table \ref{tab:median_asc} lists, for the directed dynamics with $V=0$ and $V=1$, the median
$\tau_{asc}$ of the distribution of the hitting time $t_{asc}$ of an ascending stripe state. 

These examples show trends that are also encountered for other final states (states with two vertical stripes, with one horizontal stripe, etc.).
Whereas for Glauber and $V=0$ dynamics the typical hitting time of the final state can be of the order of $30\,000$ time steps for the largest systems studied, for $V=1$ this time is never more than a few hundred time steps.
More precisely, in this latter case the typical hitting time of the final state increases
linearly with the system size $L$, while for Glauber and $V=0$ dynamics this time increases as $L^2$ (see \cite{gp2015} for a similar study of the size dependence of $\tau_g$ when $r=1$).
These features are analyzed in more detail in the next subsection.

We also note from table \ref{tab:median_10} that the typical hitting time of a stripe state is larger for $V=0$ than for the Glauber case, with larger
differences showing up for values of $r$ close to 1.

\subsection{The distribution of hitting times}
A more refined study consists in analyzing the full distribution of the hitting times.
This study is restricted here to the case $r=1$, with sizes ranging from $L=40$ to $L =160$. 

\begin{figure}[h]
\begin{center}
\includegraphics[angle=0,width=0.7\linewidth]{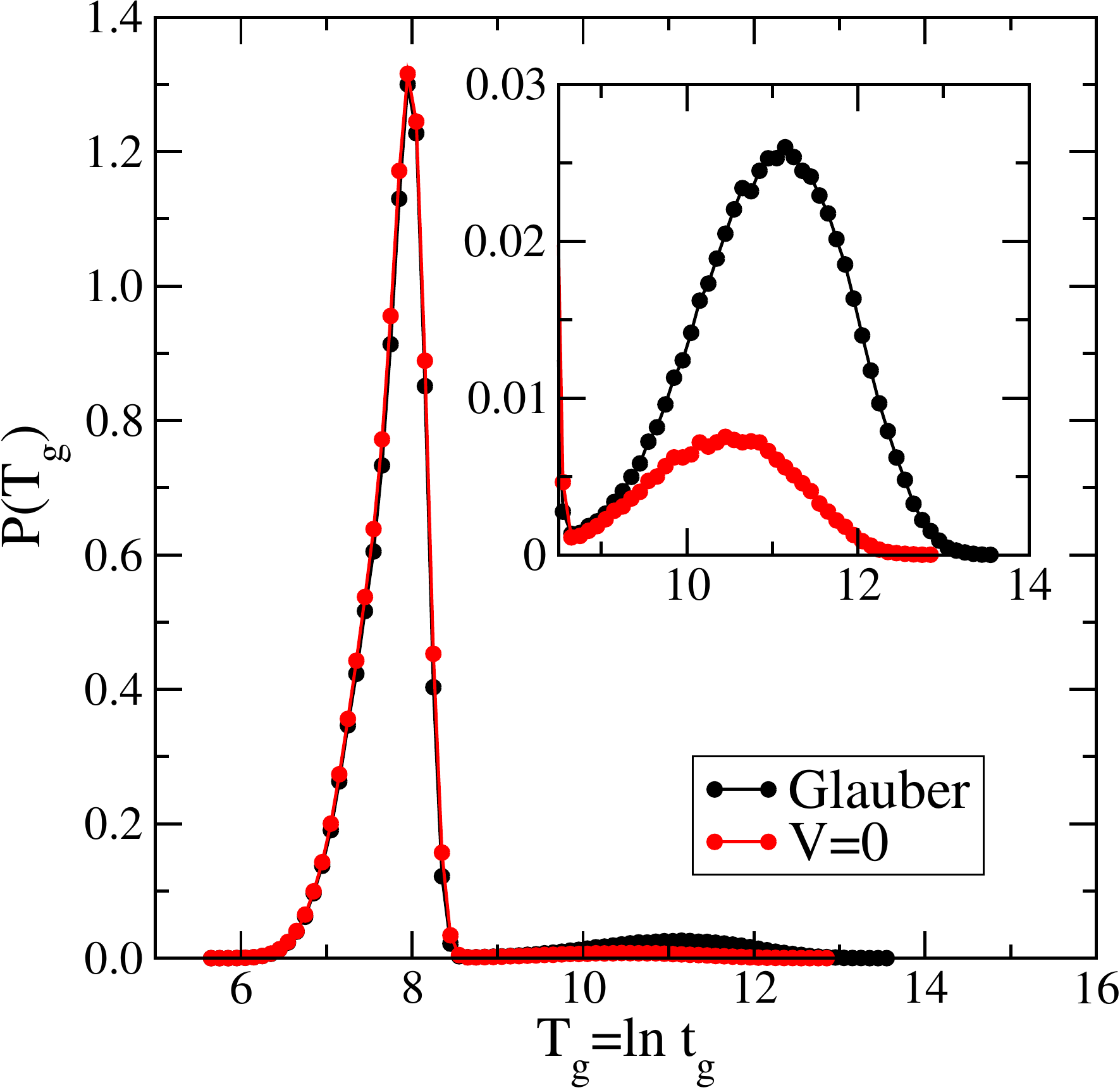}
\end{center}
\caption{Distribution of the logarithmic hitting time $T_g=\ln t_g$ of the ground state for the Glauber and $V=0$ dynamics.
The linear extent of the square system is $L=80$. 
The inset highlights the difference between the second bumps
for the two different dynamics.
}
\label{fig:Tg}
\end{figure}
\begin{figure}[h]
\begin{center}
\includegraphics[angle=0,width=0.7\linewidth]{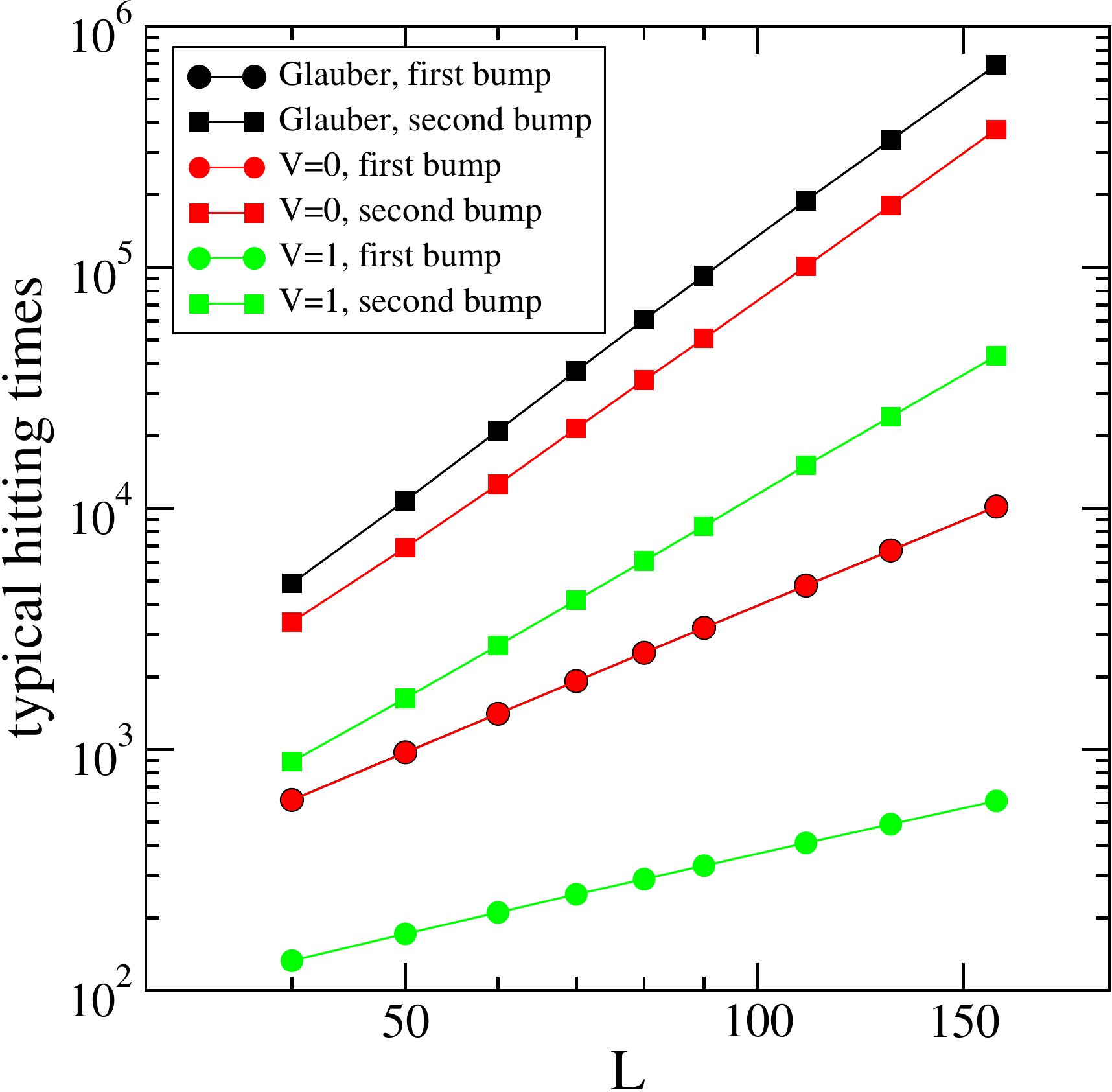}
\end{center}
\caption{Size-dependence of the typical hitting times of the ground state as measured by the
median of the times forming one the bumps (as in figure \ref{fig:Tg}). 
For the first bump $\tau_g^{(1)}\sim L^2$
for both Glauber and $V=0$ dynamics (the Glauber data are covered by those for $V=0$).
For $V=1$
a linear relationship prevails. 
For the second bump $\tau_g^{(2)}\sim L^a$ with $a$ closer to $3.5$ for both Glauber and $V=0$ dynamics, and $a$ closer to $3$ for $V=1$.
}
\label{fig:taug}
\end{figure}
\begin{figure}[h]
\begin{center}
\includegraphics[angle=0,width=0.7\linewidth]{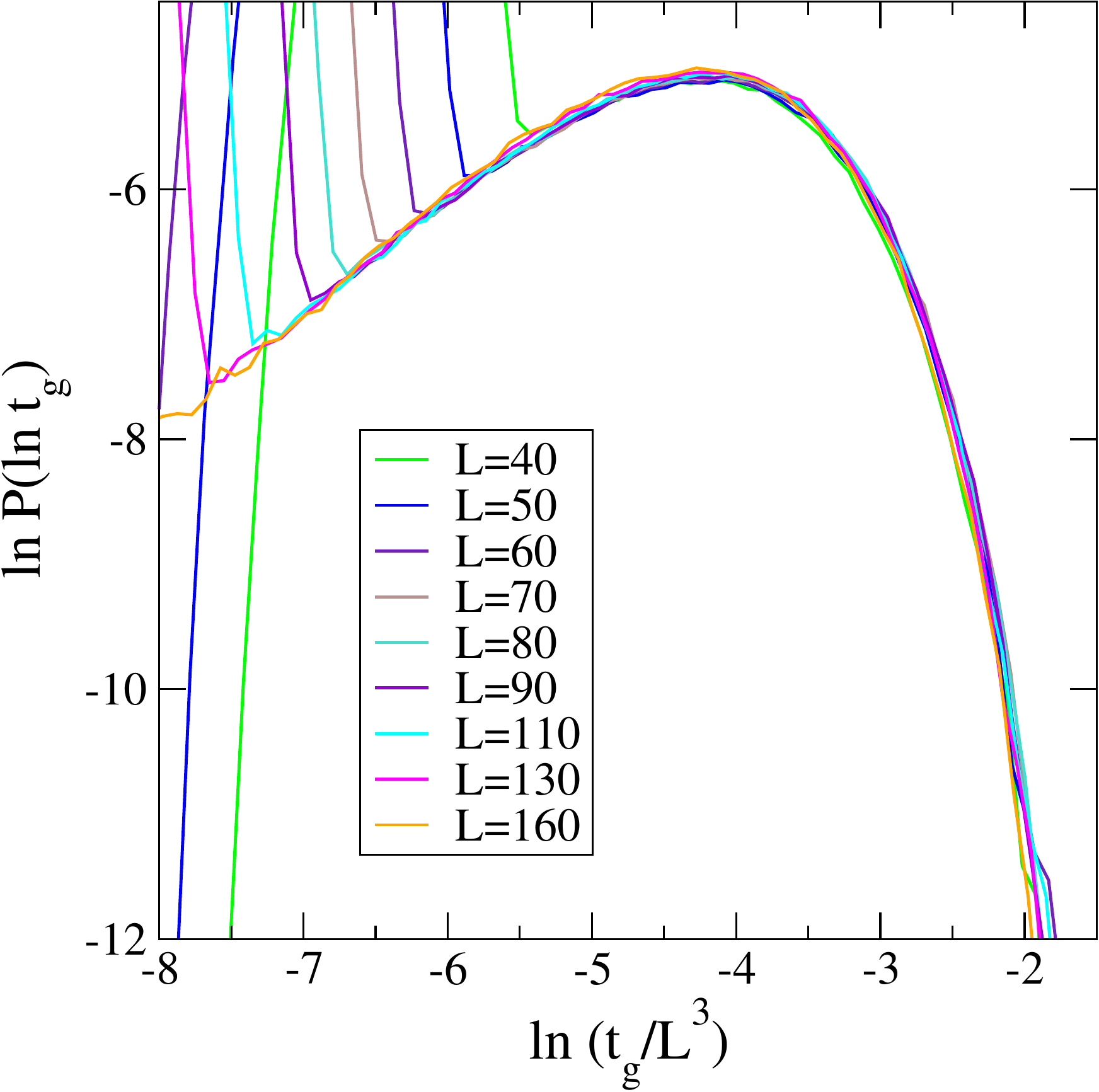}
\end{center}
\caption{Histogram of the logarithmic hitting time of the ground state $T_g=\ln t_g$ for the directed Ising model with $V=1$,
where the rescaling focuses on the second bump.
}
\label{fig:TgV1}
\end{figure}

Figure \ref{fig:Tg} depicts the distributions of the logarithmic hitting time $T_g=\ln t_g$
of the ground state, both for Glauber dynamics and $V=0$ dynamics, for a square system of size $L=80$.
These distributions are bimodal with two bumps of quite different masses, most of the histories contributing to the first one.

The typical times corresponding to the two bumps turn out to scale differently.
The first bump corresponds to histories leading to the ground state rather rapidly (on a scale of time $L^2$, see below).
This first bump is essentially the same for the two dynamics.
The second bump is due to the formation of long-lived metastable states, which are diagonal stripes disappearing at large times and correspond to a slower process (with slower typical times, see below).

This slower process is slightly different for the Glauber and $V=0$ dynamics, as can be seen in the inset of figure \ref{fig:Tg}.
In particular, for the former both ascending and descending stripes contribute, whereas for the latter only ascending ones do. 

For the directed Ising model with $V=1$ the distribution of the hitting time $t_g$ is also bimodal, where the two modes correspond again to the two different processes described above.

In order to estimate the scaling behaviour with $L$ of these two processes 
we determine the median of the times forming each of the bumps (associated to the two modes) separately, that we denote by $\tau^{(1)}_g$ and $\tau^{(2)}_g$ (see figure \ref{fig:taug}). 
For Glauber and $V=0$ dynamics $\tau^{(1)}_g\sim L^2$, as 
expected for diffusive coarsening, while for $V=1$ $\tau^{(1)}_g\sim L$, as expected for ballistic coarsening \cite{gp2015}. 
For the second bump $\tau_g^{(2)}\sim L^a$ with $a\approx 3.5$ for Glauber and $V=0$ dynamics
and $a\approx 3$ for $V=1$.
Figure \ref{fig:TgV1} depicts the histogram of the logarithmic hitting time of the ground state $T_g=\ln t_g$ for the directed Ising model with $V=1$, where the rescaling with $a=3$ focuses on the second mode.

\subsection{Back to the blocking probabilities}

The distributions of the hitting time $t_g$ contain additional information that provides 
another determination of the blocking probabilities investigated in section \ref{sec:blocking}.
We start with the Glauber case.
For each size, we count the number $n_1$ (respectively $n_2$) of histories that form the first (respectively second) bump
(as in figure \ref{fig:Tg}). 
In addition to this, we have $n_3$ histories that end in a (vertical or horizontal) stripe state. 
If the predictions of critical percolation account for the simulations of Glauber dynamics, the identification of the following ratios with the quantities defined for percolation in section \ref{sec:theory} should hold,
\begin{eqnarray}\label{eq:ratiosG}
\fl n_1/(n_1+n_2+n_3) & \hookrightarrow & \pi_{h,v}, \nonumber\\
\fl n_2/(n_1+n_2+n_3) & \hookrightarrow & \pi_d ,\nonumber\\
\fl n_3/(n_1+n_2+n_3) & \hookrightarrow & \pi_{0,1} +\pi_{1,0}.
\end{eqnarray}
Table \ref{tab:ratiosG} gives a comparison of the numerical values of these ratios for different system sizes with the corresponding theoretical values, confirming this prediction.
It is worth noting that by this method we can access separately the `hidden' probabilities $\pi_{h,v}$ and $\pi_{d}$, while the simulations described in section \ref{sec:blocking} gave only access to their sum (see \ref{eq:sumrule}).

\begin{table}[!h]
\caption{Glauber dynamics: comparison of numerical estimates of the ratios in the left side of (\ref{eq:ratiosG}) with the predictions of critical percolation theory (right side of (\ref{eq:ratiosG})).
In (\ref{eq:ratiosG}) $n_1$ and $n_2$ are the number of histories forming the first and second bumps in the distribution of the hitting time
of the ground state (as depicted for instance in figure \ref{fig:Tg} for $L=80$), while $n_3$ is the number of histories ending
in a blocked state with vertical or horizontal stripes.}
\label{tab:ratiosG}
\begin{center}
\begin{tabular}{|c|c|c|c||c|c|}
\hline
 Glauber& $L=40$ & $L=80$ & $L=160$ &percolation& \\ \hline
$n_1/(n_1+n_2+n_3)$ & 0.636 & 0.624 & 0.621 &$\pi_{h,v}$& 0.61905 \\ \hline
$n_2/(n_1+n_2+n_3)$ & 0.027 & 0.038 & 0.040 &$\pi_{d}$& 0.04212 \\ \hline
$n_3/(n_1+n_2+n_3)$ & 0.337 & 0.339 & 0.339 &$\pi_{0,1}+\pi_{1,0}$& 0.33882 \\
\hline
\end{tabular}
\end{center}
\end{table}

Turning now to the case of the directed Ising model, in addition to $n_1, n_2$ and $n_3$ defined as above, the number $n_4$ of histories leading to an ascending stripe state should also be considered.
In order to test the validity of the predictions of percolation we have to compare the ratios in the left side of the following equation to the quantities in the right side, defined in section \ref{sec:theory}:
\begin{eqnarray}\label{eq:ratiosV0}
\fl n_1/(n_1+n_2+n_3+n_4) & \hookrightarrow & \pi_{h,v}, \nonumber\\
\fl n_2/(n_1+n_2+n_3+n_4) & \hookrightarrow & \pi_{desc} ,\nonumber\\
\fl n_3/(n_1+n_2+n_3+n_4) & \hookrightarrow & \pi_{0,1} +\pi_{1,0},\nonumber\\
\fl n_4/(n_1+n_2+n_3+n_4) & \hookrightarrow & \pi_{asc}\,.
\end{eqnarray}
Table \ref{tab:ratiosV0} gives a comparison of the numerical values of these ratios for different system sizes with the corresponding theoretical values.
These results
confirm the discrepancies observed in section \ref{sec:blocking} between the dynamical quantities obtained by simulation and the prediction of percolation theory.

\begin{table}[!h]
\caption{Directed Ising model: comparison of numerical estimates of the ratios in the left side of (\ref{eq:ratiosV0}) with the predictions of critical percolation (right side of (\ref{eq:ratiosV0})).
In (\ref{eq:ratiosV0}) $n_1$ and $n_2$ are the number of histories forming the first and second bumps in the distribution of the hitting time
of the ground state (as depicted for instance in figure \ref{fig:Tg} for $L=80$), $n_3$ is the number of histories ending
in a (vertical or horizontal) stripe state and $n_4$ is the number of histories leading to an ascending stripe state.}
\label{tab:ratiosV0}
\begin{center}
\begin{tabular}{|c|c|c|c||c|c|}
\hline
$V=0$ & $L=40$ & $L=80$ & $L=160$ & percolation& \\ \hline
$n_1/(n_1+n_2+n_3+n_4)$ & 0.630 & 0.624 & 0.623 &$\pi_{h,v}$& 0.61905 \\ \hline
$n_2/(n_1+n_2+n_3+n_4)$ & 0.007 & 0.010 & 0.010 &$\pi_{desc}$& 0.02106 \\ \hline
$n_3/(n_1+n_2+n_3+n_4)$ & 0.330 & 0.331 & 0.331 &$\pi_{0,1}+\pi_{1,0}$& 0.33882 \\ \hline
$n_4/(n_1+n_2+n_3+n_4)$ & 0.033 & 0.035 & 0.036 &$\pi_{asc}$& 0.02106 \\ \hline
$V=1$ & $L=40$ & $L=80$ & $L=160$ & percolation &\\ \hline
$n_1/(n_1+n_2+n_3+n_4)$ & 0.629 & 0.624 & 0.622 &$\pi_{h,v}$& 0.61905 \\ \hline
$n_2/(n_1+n_2+n_3+n_4)$ & 0.0075 & 0.0097 & 0.0106 &$\pi_{desc}$& 0.02106 \\ \hline
$n_3/(n_1+n_2+n_3+n_4)$ & 0.330 & 0.331 & 0.331 &$\pi_{0,1}+\pi_{1,0}$& 0.33882 \\ \hline
$n_4/(n_1+n_2+n_3+n_4)$ & 0.033 & 0.035 & 0.036 &$\pi_{asc}$& 0.02106 \\
\hline
\end{tabular}
\end{center}
\end{table}

\subsection*{Remark}
A natural quantity to consider is the complementary distribution function $\prob(t_g>t)$ (or $\prob(T_g>T)$, where $T=\ln t$).
A similar quantity is investigated in \cite{spirin1,olejarz} for Glauber dynamics, namely the survival probability $S(t)$, defined as the probability that there still exist flippable spins in the system.

\section{Discussion}

The present study comes after many previous investigations on similar questions \cite{spirin1,spirin2,barros,olejarz,picco}.
It  completes these works and departs from them in several ways.

For Glauber dynamics, the final states are determined until cessation of activity.
The distribution of the hitting time $t_g$ (or of the logarithmic time $T_g=\ln t_g$) provides a separate determination of the probabilities $\pi_{h,v}$ and $\pi_{d}$, which are otherwise inaccessible directly\footnote{These probabilities can also be determined by the analysis of the short-time regime \cite{private}.}.

The main purpose of the present work is nevertheless the comparison of the properties of the directed Ising model with those of Glauber dynamics with regard to the fate of system of spins on a square or rectangular lattice, evolving at zero temperature from a random disordered initial condition, and freezing either in the ground state or into infinitely long-lived metastable states.

The main observation made here is that, while there is convincing evidence that the predictions of critical percolation coincide with the measured values of the blocking probabilities by simulations of Glauber dynamics, 
as has been already demonstrated previously in \cite{barros,olejarz,picco}, 
this is no longer the case for the directed Ising model.
Assuming the claim made in \cite{olejarz} that critical percolation predicts the values of the blocking probabilities for any curvature-driven non-conserved dynamics, one infers that
conversely, since the predictions of percolation are not entirely satisfied for the directed Ising model, then the origin of these discrepancies 
should lie in the lack of isotropy of the dynamics of this model.
For example, as shown in \cite{gp2015,gl2017}, a minority square does not melt isotropically.

\begin{figure}[h]
\begin{center}
\includegraphics[angle=0,width=0.7\linewidth]{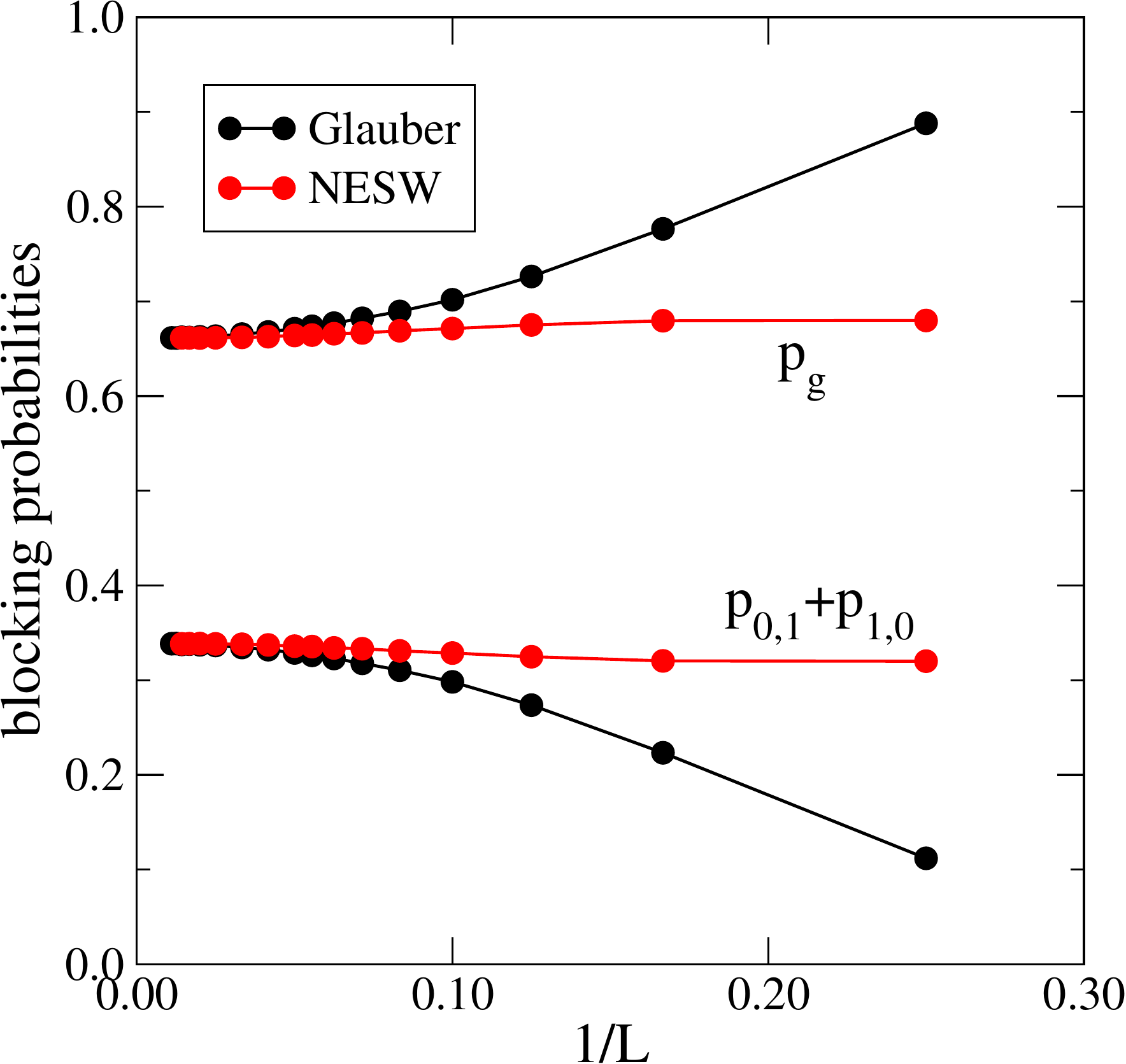}
\end{center}
\caption{Blocking probabilities for Glauber dynamics and the NESW model.
}
\label{fig:NESW}
\end{figure}

A way of completing the picture consists in considering another zero-temperature dynamics, where the four cardinal directions play the same role, defined by the following rate
\beqa\label{eq:rate4}
w=\frac{\alpha}{2\times 4}
\Big[&(&1- \s\s_E)(1- \s\s_N)
+(1- \s\s_W)(1- \s\s_S)+
\nonumber\\
&(&1- \s\s_N)(1- \s\s_W)
+(1- \s\s_S)(1- \s\s_E)
\Big].
\eeqa
This NESW model  is now isotropic:
the four corners of a minority square are equally eroded by the dynamics.
It does not generate ascending or descending stripe states.
Note however that, since the two last configurations of figure \ref{fig:movesGlau} have zero rate, this dynamics
does not depend on $\Delta E$ only.
The question is now whether the blocking probabilities $p_g$ or $p_{0,1}$ for this model are the same as for Glauber dynamics.
Simulations show that, while this does not hold for a finite size $L$, asymptotically, for $L\to\infty$, the blocking probabilities are the same for the two models as depicted in figure \ref{fig:NESW}.

To close, let us note that the blocking probabilities for the directed Ising model seems independent of the value of the asymmetry parameter $V$ (this has been confirmed by the investigation of other values of $V$).
In some sense this feature is a remnant of the property of the model where at any finite temperature observables are independent of $V$ at stationarity.
In contrast, the hitting times, which are are dynamical quantities, do depend on $V$. 
The scaling of these hitting times is quite different whether $V=0$ (reversible dynamics) or $V=1$, or more generally $V\ne0$ (irreversible dynamics).
For the latter case there is acceleration of the dynamics, a feature already observed in \cite{gp2015}, and of current interest \cite{jack}.

\ack
We wish to thank J-M Luck for many useful conversations and M Picco for sharing with us preliminary results on the directed Ising model obtained by the methods of \cite{picco}, and for a careful reading of our manuscript.
MP is supported by the US National Science Foundation through grant DMR-1606814.

\appendix

\section{Rules of the dynamics for the directed Ising model}
\label{sec:def}

In this appendix we give the general definition of the dynamical rules of the kinetic Ising model with asymmetric dynamics, named for short the {\it directed (kinetic) Ising model}, valid for any temperature. 
We borrow the material of this section from \cite{gp2015}.

\subsection{Expression of the rate}

We consider a system of Ising spins on the square lattice
of linear size $L$, with periodic boundary conditions.
The Hamiltonian is given in (\ref{eq:Ising}).
We set $J=1$, $k_B=1$ and denote the reduced coupling constant by $K=1/T$.
At each instant of time, a spin, denoted by $\s$, is chosen at random and flipped with rate $w$.
We choose the following form of the rate \cite{gb2009,gp2014,cg2013}, with the notation $\{\s_E,\s_N,\s_W,\s_S\}$ for the East, North, etc., neighbours of the central spin $\s$, 
\beq\label{eq:rate}
\fl w=\frac{\alpha}{2}
\left[\frac{1+V}{2}(1- \gamma\s\s_E)(1- \gamma\s\s_N)
+\frac{1-V}{2}(1- \gamma\s\s_W)(1- \gamma\s\s_S)\right],
\eeq
where $\alpha$ fixes the scale of time, $\gamma=\tanh 2K$ and $V$ is the asymmetry (or irreversibility) parameter, which allows to interpolate between the symmetric case ($V=0$) and the totally asymmetric ones ($V=\pm1$).
This expression of the rate satisfies the condition of global balance, that is to say, leads to a Gibbsian stationary measure with respect to the Hamiltonian (\ref{eq:Ising}) even if the condition of detailed balance is not satisfied\footnote{We refer the reader to \cite{gb2009,cg2013} for a detailed account of this property.}.
It represents one, among many, possible expression of a rate, for the kinetic Ising model on the square lattice, possessing this property.
This expression is invariant under up-down spin symmetry.
In (\ref{eq:rate}), for $V>0$ (resp. $V<0$) the couple (North-East) (resp. (South-West)) is more influential on the central spin than the other one. 

Here we focus our attention on the two particular cases of symmetric dynamics ($V=0$), and completely asymmetric dynamics ($V=\pm1$).
First, if $V=0$, the rate
\beq
w=\frac{\alpha}{2}\big[
1-\frac{\gamma}{2}\s(\s_E+\s_N+\s_W+\s_S)
+\frac{\gamma^2}{2}(\s_E\s_N+\s_W\s_S)
\big]
\label{eq:rate0}
\eeq
satisfies the condition of detailed balance \cite{gb2009,cg2013}.
In constrast with the Glauber rate
\beq\label{glau2D}
w=\frac{\alpha}{2}\big[1-\s\tanh [K(\s_E+\s_N+\s_W+\s_S)]\big],
\eeq
which is fully symmetric under a permutation of the neighbouring spins, or, equivalently, only depends on the variation of the energy due to a flip, 
\beq
\Delta E=2\sigma (\sigma_E+\sigma_N+\sigma_W+\sigma_S),
\eeq
the rate (\ref{eq:rate0}) with $V=0$ is not fully symmetric under a permutation of the neighbouring spins and does not depend on $\Delta E$ only.
For instance, if the local field 
\beq
h=\s_E+\s_N+\s_W+\s_S
\eeq
vanishes then the rate (\ref{eq:rate0}) takes two different values, $\alpha(1\pm\gamma^2)$, according to the configuration of the neighbours $\{\s_E,\s_N,\s_W,\s_S\}$: $\alpha(1+\gamma^2)$ corresponds to $\{++--\}$ or $\{--++\}$, while $\alpha(1-\gamma^2)$ corresponds to all other configurations.
This is a manifestation of an anisotropy in the dynamics.
By comparison, for the Glauber case, if the local field vanishes, the rate takes only one value, $\alpha/2$.

The case with $V=1$ corresponds to the totally asymmetric dynamics where the central spin $\s$ is influenced by its East and North neighbours only.
Then
\beq\label{eq:ne}
w=\frac{\alpha}{2}
\left[1-\gamma\s(\s_E+\s_N)+\gamma^2\,\s_E\s_N\right].
\eeq
A similar expression, involving $\s_W$ and $\s_S$, holds for $V=-1$:
\beq\label{eq:sw}
w=\frac{\alpha}{2}
\left[1-\gamma\s(\s_W+\s_S)+\gamma^2\,\s_W\s_S\right].
\eeq
A remarkable fact about these expressions is that they are {\it unique}, up to the scale of time fixed by the choice of $\alpha$, in the following sense.
For rates only involving NEC (North, East, Central) spins, or SWC (South, West, Central) spins,
i.e., when only two neighbours, chosen among the four possible ones, are influential upon the central spin,
imposing the condition of global balance uniquely determines the expressions (\ref{eq:ne}) or (\ref{eq:sw}) \cite{gb2009,cg2013}.
The rate (\ref{eq:ne}) for the totally asymmetric NEC dynamics was originally given in \cite{kun}, under a slightly different form, but without considerations on its derivation or its unicity.
Alternatively, instead of NEC or SWC spins, the choice of NWC or  SEC ones is equally possible.

Let us add a comment on terminology.
We refer to the kinetic Ising model defined by the rate (\ref{eq:rate}) as the {\it directed Ising model}, as done in previous works \cite{gb2009,gp2014,gp2015}, even for the more symmetrical case $V=0$ (with rate (\ref{eq:rate0})), where dynamics is reversible.

\subsection{Choice of the scale of time}

Since in the course of this work we systematically compare the results obtained with the rate (\ref{eq:rate}) to those obtained with Glauber rate (\ref{glau2D}), we now discuss the choice of the time-scale parameter $\alpha$ made in our simulations.

At infinite temperature ($\gamma=0$) the two expressions (\ref{eq:rate}) and (\ref{glau2D}) yield rates equal to $\alpha/2$,
hence the two time scales of these dynamics are identical.
On the other hand, in a simulation at finite (or zero) temperature, if, for practical purposes, one wants the rates to be less than 1, one should fix the value of $\alpha$ according to the largest rate, {\it simultaneously for both rates}.
The largest values taken by these rates are
\beqa
{\rm Glauber}\,: \frac{\alpha}{2}(1 +\gamma)^2/(1 +\gamma^2) &&\zeq{T\to0} \frac{\alpha}{2}\times 2,
\nonumber\\
{\rm Directed}:\frac{\alpha}{2}(1+\gamma)^2 &&\zeq{T\to0} \frac{\alpha}{2}\times 4.
\eeqa
In the simulations presented in the bulk of the paper the common choice $\alpha=1/2$ has been done for both expressions.

\end{document}